\documentclass[11pt,a4paper,twoside]{article}
\usepackage[utf8]{inputenc}
\usepackage[T1]{fontenc}
\usepackage{amsmath,amsfonts,amssymb}
\usepackage{graphicx}
\usepackage{booktabs}
\usepackage{multirow}
\usepackage{float}
\usepackage{caption}
\usepackage[colorlinks=true,linkcolor=blue,citecolor=blue]{hyperref}
\usepackage[numbers]{natbib}
\usepackage{afterpage}
\usepackage{authblk}
\usepackage{geometry}
\graphicspath{ {./tmp/} }

\usepackage{fancyhdr}
\fancypagestyle{plain}{
	\fancyhf{}
	
}
\title{\textbf{CiUNet: A Hybrid Swin-CNN UNet\\ for Medical Image Segmentation}}

\author{Bin Dong, Jinghong Chen*}
\affil{\small{Ciphowork GmbH} \\ \small{b.dong@ciphowork.de, jh.chen@ciphowork.de}}
\date{}

\begin{document}

	\maketitle
	
	\begin{abstract}
		Medical image segmentation requires high accuracy and robustness, yet practical commercial deployment also demands privacy preservation and computational efficiency. In this context, the U-Net architecture, which can be inherently decoupled into independent encoder and decoder components, serves as a natural commercial choice. However, pure Transformer-based variants like Swin-UNet often suffer from insufficient local detail capture and limited interpretability. In this paper, we propose a lightweight hybrid architecture built upon the Swin-UNet framework. Our model integrates a parallel CNN encoder to complement the shallow layer reasoning of Swin Transformers with local texture features. To bridge the semantic gap and enhance fine-grained spatial detail recovery, we design an asymmetric feature fusion strategy and introduce cross-layer skip (XSkip) connections that explicitly propagate shallow CNN features into the decoder. We further incorporate novel loss functions and an auxiliary supervision head (Aux-Head) to strengthen training stability, boundary delineation, and intermediate feature interpretability. Extensive experiments on the Synapse multi-organ segmentation dataset demonstrate that our approach achieves state-of-the-art competitive Dice scores and Hausdorff distances, offering an accurate, efficient, and interpretable solution for clinical deployment. Due to our commercial considerations, we publicly release only the trained weights on the Synapse dataset and a lightweight demo predictor at \url{https://github.com/ciphoBD/CiUNet}.
	\end{abstract}
	
	\section{Introduction}
	Medical image segmentation is a fundamental task in clinical workflows, facilitating disease diagnosis, surgical navigation, and treatment assessment. For decades, Convolutional Neural Networks (CNNs), especially the U-Net architecture \cite{ronneberger2015u}, have served as the de facto standard for this task. The U-Net’s encoder-decoder structure with skip connections effectively integrates semantic context with spatial details. However, the intrinsic locality of convolutional operations restricts CNNs from capturing long-range global dependencies, often leading to sub-optimal boundaries for anatomically complex organs. Following these advances, Vision Transformers (ViTs) \cite{dosovitskiy2020vit} have emerged as a powerful alternative to CNNs due to their self-attention mechanism, which excels at modeling long-range dependencies \cite{vaswani2017attention}. Building upon this, TransUNet \cite{chen2021transunet} and Swin-UNet \cite{cao2021swinunet} introduce the Transformer as a strong encoder for medical image segmentation. In particular, Swin-UNet inherits the hierarchical Swin Transformer blocks with shifted window attention, achieving superior performance by reducing the quadratic complexity of standard ViTs.
	Despite these advancements, Swin-UNet still suffers from three critical limitations stemming from its pure Transformer design: insufficient preservation of local textural details due to the lack of convolutional inductive bias; inaccurate spatial reconstruction caused by inadequate feature propagation across skip connections; and limited interpretability and training instability in its deep decoder architecture. This work attempts to mitigate these deficiencies by revisiting the encoder-decoder paradigm of Swin-UNet, with specific enhancements tailored to each of the above bottlenecks.
	
	To systematically validate our architectural improvements, we conduct our evaluation on the Synapse multi-organ CT dataset \cite{landman2015synapse}, a widely used benchmark for abdominal organ segmentation. This dataset consists of 30 contrast-enhanced CT scans comprising 3,779 axial slices, with ground-truth annotations for eight distinct abdominal organs: the aorta, gallbladder, spleen, left/right kidney, liver, pancreas, and stomach. We select the Synapse dataset because it features diverse organ sizes, highly variable patient anatomies, and clinically complex boundaries—particularly for the pancreas and gallbladder—which makes it a rigorous testbed for evaluating advanced segmentation frameworks. To comprehensively assess segmentation performance, we employ the Dice Similarity Coefficient (DSC) for measuring region-wise volumetric overlap, and the 95\% Hausdorff Distance (HD95) as a boundary-aware metric. While DSC reflects the overall segmentation quality, HD95 is clinically critical as it reveals the largest boundary misalignment, which directly impacts downstream tasks such as surgical navigation and radiation therapy planning.
	
	Trained and rigorously evaluated on the Synapse dataset, we introduce \textbf{CiUNet}—a novel hybrid U-shaped architecture specifically designed to overcome the above limitations of Swin-UNet. Our contributions are three-fold:
	\begin{enumerate}
		\item \textbf{Asymmetric Dual-Encoder Framework Upgrades}: Instead of a single Swin encoder, we introduce a parallel lightweight ResNet encoder to explicitly capture local spatial details. The Swin branch remains the primary backbone for global context, while the ResNet branch provides complementary high-resolution features.
		
		\item \textbf{XSkip Mechanism for Decoder Enhancement}: To recover lost spatial resolution, we replace standard skip connections with our proposed XSkip module. This mechanism selectively aligns and propagates shallow CNN features into deeper decoder layers via spatial attention, ensuring sharper boundary delineation.

		\item \textbf{Multi-Level Auxiliary Supervision}: To improve the interpretability and training stability of the deep Swin decoder, we introduce auxiliary segmentation heads. By providing multi-level gradient guidance, this supervision aligns intermediate features with clinical expectations, significantly accelerating convergence and mitigating the black-box nature of the Transformer backbone.
	\end{enumerate}
	
	Complementing these technical improvements, the inherent encoder-decoder separation of UNet architecture offers a strategic commercial security advantage. Sensitive patient CT images can be processed locally at the lightweight encoder layer, with only high-level semantic feature vectors transmitted to the cloud for the heavy decoder computation. This decoupling enables a privacy-preserving SaaS (Software-as-a-Service) model without compromising inference accuracy, making secure hospital-cloud collaboration feasible.
	
	Finally, due to current commercial deployment plans, we are unable to release the full implementation or the instance-level inference heads at this time. To support reproducibility and practical evaluation, we provide a light-weight ONNX runtime predictor with a desktop GUI, operating at a fixed $224 \times 224$ resolution. The model achieves a competitive mean Dice score of 83.72\% on the Synapse dataset with only 31.68M trainable parameters, demonstrating that our framework provides an accurate, efficient, and commercially viable solution for medical image segmentation.
	\clearpage
	
	\section{Related Work}
	
	\subsection*{Medical Image Segmentation}
	Medical image segmentation is a cornerstone of modern computer-aided diagnosis, requiring accurate voxel-level or pixel-level delineation of anatomical structures from modalities such as CT and MRI. Over the past decade, deep learning has fundamentally reshaped this field, progressing from purely convolutional architectures to advanced hybrid designs.
	
	\begin{figure}[htbp]
		\centering
		\includegraphics[width=1.0\linewidth, keepaspectratio]{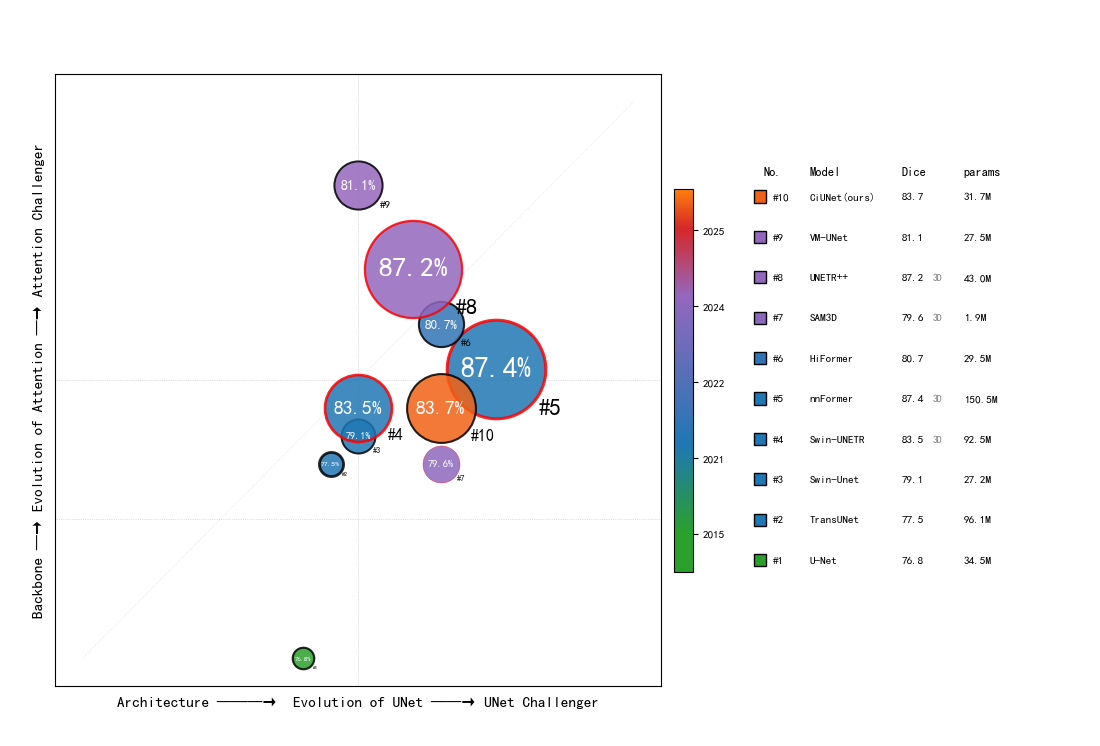}
		\caption{Comparison of our CiUNet with other methods in terms of parameter count and DSC on the Synapse dataset.}
		\label{fig:results}
	\end{figure}
	
	\subsection*{2D Convolution-based Segmentation Methods}
	The U-Net architecture \cite{ronneberger2015u} established the de facto standard for 2D medical image segmentation. Its symmetric encoder-decoder structure, augmented with long-range skip connections, effectively integrates high-level semantic features with fine-grained spatial details. Subsequent refinements, including Res-UNet \cite{drozdzal2016resunet} for deeper feature propagation, U-Net++ \cite{zhou2018unet++} with nested dense skip pathways, and UNet3+ \cite{huang2020unet} with full-scale skip connections, have demonstrated that architectural optimization of the U-shaped backbone can yield substantial performance gains. Despite these advancements, the intrinsic locality of convolution operations fundamentally restricts CNNs from modeling global long-range dependencies, which are essential for understanding the complete anatomical topology of complex organs.
	
	\subsection*{2D Transformer-based Segmentation Methods}
	The emergence of Vision Transformers (ViTs) \cite{dosovitskiy2020vit} offered a paradigm shift by introducing global self-attention mechanisms originally developed for Natural Language Processing (NLP) into computer vision. The Swin Transformer \cite{liu2021swinv1} improves upon ViT’s quadratic complexity through a hierarchical architecture and shifted window attention, making Transformers computationally viable for dense prediction tasks. In medical imaging, Swin-UNet leverages the Swin Transformer as a pure encoder-decoder backbone, achieving superior performance by directly capturing long-range semantic correlations. However, pure Transformer architectures often suffer from data hunger, requiring extensive pre-training to generalize effectively. More critically, the lack of CNN's inherent inductive bias leads to poor preservation of low-level textures, resulting in blurred boundaries and inaccurate spatial localization in segmentation outputs.
	
	\subsection*{2D Hybrid Segmentation Methods}
	To reconcile the global reasoning of Transformers with the local precision of CNNs, hybrid approaches have gained substantial traction. TransUNet \cite{chen2021transunet} pioneered this direction by utilizing a CNN for initial feature extraction and a Transformer for global context encoding. HiFormer \cite{heidari2023hiformer} introduced a double-branch encoder combined with a Double-Level Fusion (DLF) module to seamlessly merge Swin Transformer and CNN features. MISSFormer \cite{huang2021missformer} proposed an Enhanced Transformer Block and cross-scale context bridges to improve discriminative feature learning. VM-UNet \cite{ruan2024vm} further explored the potential of pure state-space models for medical segmentation, demonstrating the growing diversity of backbone designs. Despite these advances, many existing hybrid frameworks either rely on naive concatenation mechanisms that fail to maintain feature consistency across scales, or introduce complex fusion modules that significantly increase computational overhead.
	
	\subsection*{3D Segmentation Methods}
	Beyond 2D slice-based approaches, 3D segmentation methods have gained prominence for their ability to exploit volumetric consistency across consecutive slices. Unlike 2D models that operate on isolated frames, 3D architectures—such as 3D U-Net \cite{iek2016unet3d} and V-Net \cite{milletari2016vnet} —utilize full-organ volumetric information to improve semantic coherence and reduce inter-slice discontinuity. These methods are generally considered a necessary means of improving practical segmentation robustness, particularly for organs with complex 3D structures, as they leverage spatial continuity to suppress noise and refine organ boundaries. However, 3D segmentation is often regarded as a supplementary strategy in clinical workflows due to its substantial computational and memory demands. The trade-off between the improved volumetric consistency and the increased inference latency must be carefully managed. In this context, our work remains focused on the 2D domain but incorporates an innovative architectural design that mimics 3D-like context reasoning at a fraction of the computational cost, making it more suitable for real-time deployment and privacy-preserving commercial applications.

	\section{Method}
	
	\subsection{Architecture Overview}
	The overall architecture of our proposed CiUNet is illustrated in Fig.\ref{fig:architecture}. It is a dual-encoder U-shaped network designed for 2D medical image segmentation. The framework comprises a Swin Transformer encoder, a CNN encoder, a feature fusion module, a decoder with additional cross-layer skip connections, and auxiliary supervision heads. Specifically, compared to the original Swin-Unet \cite{cao2021swinunet} which organizes the encoder into shallow (Layer1), mid (Layer2), deep (Layer3), and bottleneck (LayerBN) layers, our architecture introduces an additional high-resolution Layer0 at the very front. In this backbone, the Swin Transformer encoder operates from Layer1 to LayerBN, while the parallel CNN encoder is tailored to extract local features from Layer0 to Layer3. Notably, to leverage the inherent strengths of the pre-trained visual priors, both the Swin and CNN branches freeze the feature extraction modules of the initial layers, restricting trainable parameters to the subsequent down-sampling layers. To handle the dual-encoder outputs, we implement a dedicated fusion module that integrates features from the same resolution layer of both branches. The fusion mechanism operates dynamically: when only a single branch's feature is available, it is directly propagated; whereas, when both are present, the Swin feature serves as the dominant semantic representation, with the CNN feature acting as a complementary spatial refinement. For the upsampling path, unlike the standard skip connections in original Swin-Unet \cite{cao2021swinunet}, we augment the decoder with additional cross-layer skip connections. These connections directly propagate the higher-resolution fine-grained details from Layer0 and Layer1 of the encoders to the corresponding decoder layers, effectively supplementing the semantic information omitted during the down-sampling process. Regarding the core attention mechanism,we adopt the LogSpacedCPB module from Swin Transformer V2 \cite{liu2021swinv2} to replace the conventional relative position bias used in standard Swin Transformer blocks. Finally, we incorporate auxiliary supervision heads (Aux-Head) attached to the intermediate decoder outputs. Benefiting from the incorporation of CNN features, the feature representations in deeper layers gain enhanced interpretability. By propagating the gradients through the auxiliary losses as well, we effectively accelerate the convergence of the network.

	\vfill
	\begin{center}
		\includegraphics[height=\dimexpr\pagegoal-\pagetotal-1.5cm\relax, keepaspectratio]{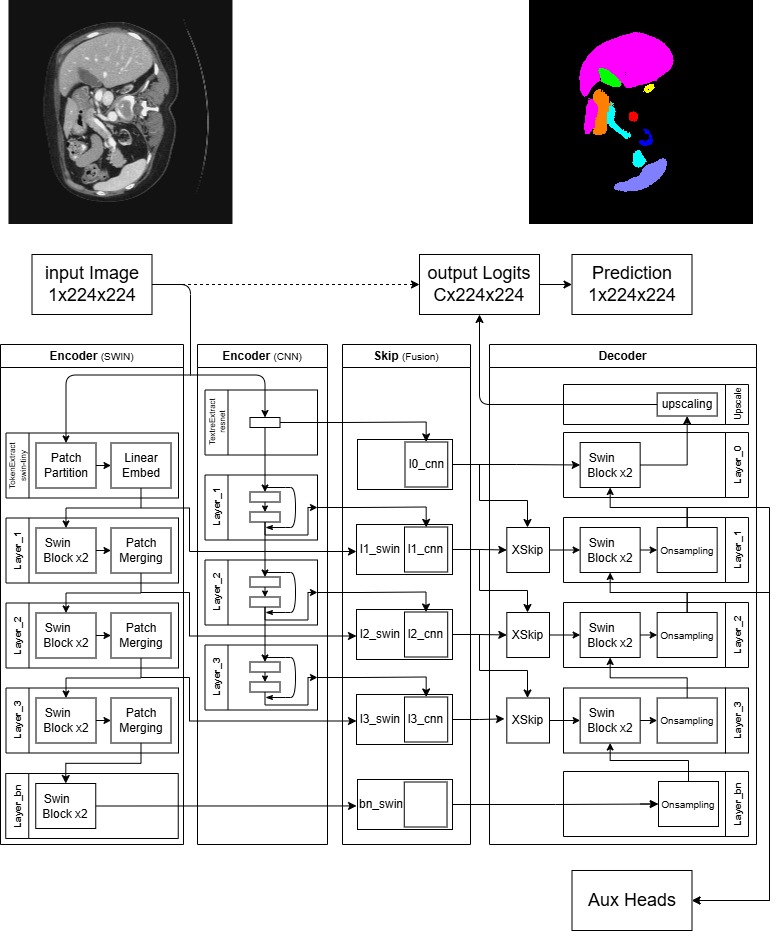}
		\captionof{figure}{The architecture of CiUNet}
		\label{fig:architecture}
	\end{center}
	\vfill
	\clearpage

	\subsection{Swin Transformer Block}
	The backbone employs the Swin Transformer block as its core attention unit, initialized using the Swin-Tiny pre-trained weights. Its fundamental principle relies on partitioning feature maps into non-overlapping windows for local self-attention and employing shifted windows (SW-MSA) to enable cross-window interactions, achieving long-range dependency modeling with linear computational complexity. 
	While we do not fully migrate to the complete Swin Transformer V2 architecture \cite{liu2021swinv2}, we adopt one of its key innovations---the Log-Spaced Continuous Positional Bias (LogSpacedCPB)---to replace the original fixed-size learnable look-up table. Instead of discrete coordinate indexing, a lightweight MLP directly regresses the bias from continuous relative coordinates \((\delta x,\delta y)\). Crucially, these raw coordinates undergo a non-linear transformation: \(\operatorname{sign}(x) \cdot \log(1+|x|)\) before being fed into the MLP. This continuous refinement eliminates resolution constraints, enabling the model to extrapolate positional biases for unseen coordinate ranges and ensuring robust generalization across varying input scales, a capability that inherently benefits from the design philosophy of Swin Transformer V2.
	
	\subsection{Dual-Encoder}
	To simultaneously capture global context and local spatial details, our framework employs a dual-encoder architecture comprising a Swin Transformer encoder and a parallel CNN encoder. The Swin encoder, initialized with Swin-Tiny pre-trained weights, prioritizes global semantic reasoning, while the CNN encoder, leveraging a pre-trained ResNet backbone \cite{he2016deep}, focuses on retaining high-resolution local textures. 
	Both encoders operate with frozen feature extraction. Specifically, the output of the shallow Swin layer (Layer1) and the output of the surficial CNN layer (Layer0) are directly designated as the frozen texture extraction results. This design preserves the general visual priors inherited from the pre-trained weights while restricting the trainable parameters to the deeper down-sampling layers. 
	Unlike TransUNet \cite{chen2021transunet}, which utilizes a CNN backbone as a primary encoder for initial feature extraction and feeds the output into a Transformer, our CNN branch is explicitly designed as a parallel encoder. In our framework, the CNN features only receive higher weight contributions during the shallow and mid layer (Layer1 and Layer2) feature fusion, acting as complementary local cues to the dominant Swin encoder.
	
	\subsection{Fusion for Skip and Bottleneck}
	Prior to the decoding phase, the network must prepare multi-scale skip features and the deepest semantic bottleneck representation. To fuse these dual-branch features, we isolate a dedicated Fusion module. As implemented in our code, the fusion mechanism operates dynamically depending on the availability of features from the Swin and CNN encoders. At each resolution layer (from Layer0 down to the LayerBN), the module first checks the input features. If both encoder branches provide valid features, they are concatenated along the channel dimension and passed through a lightweight fusion network composed of a linear projection, LayerNorm, and GELU activation, projecting the combined channels to a unified target dimension. If only a single branch's feature is present, the module directly propagates it via an identity mapping. Finally, the module outputs the fused bottleneck feature to the decoder's first upsampling layer, along with the fused skip connections for the subsequent layers. Notably, the complementary design of the dual encoders leads to asymmetric fusion patterns at the extremes of the hierarchy: the bottleneck (LayerBN) is exclusively derived from the Swin branch, as the CNN encoder does not extend to this deepest semantic layer to avoid excessive depth; conversely, the Layere0 features are solely provided by the CNN encoder, as the Swin branch's frozen shallow texture extraction is reserved for global context aggregation at later layers. This targeted feature availability ensures that the fusion module effectively aggregates global semantics from the Swin branch and high-resolution local details from the CNN branch without redundancy.

	\subsection{Decoder}
	The decoder adopts a hierarchical architecture to progressively recover the spatial resolution of the feature maps. Unlike the original Swin-Unet \cite{cao2021swinunet}, which adopts a strictly symmetric decoder mirrored from the encoder, our decoder is constructed asymmetrically to account for the dual-encoder design. The framework consists of four successive layers. At the first layer (LayerBN), a single dedicated upsampling layer is used. For the subsequent three layers (Layer3 to Layer1), each consists of two consecutive Swin Transformer blocks \cite{liu2021swinv1} with the standard configuration (window sizes and shift sizes kept unchanged from the original implementation), followed by upsampling operations.
	To upsample the deep features, we introduce a refined upsampling operation that replaces the standard patch expansion used in the original Swin-Unet. Instead of simple spatial rearrangement, this operation jointly utilizes bilinear interpolation and a lightweight dynamic weighting mechanism. This ensures smoother resolution recovery and effectively improves the reconstruction quality of the segmentation maps. Parallel to the upsampling operations, the skip connections are further enhanced with two key components. First, to bridge the semantic gap between shallow and deep feature maps, the high-resolution shallow features are spatially aligned and channel-compressed via bilinear interpolation and $1 \times 1$ convolutions before being fused with the deep decoder features. Second, a spatial attention mechanism is integrated into each skip connection pathway. Prior to the final channel concatenation, this module adaptively recalibrates the skip features by computing spatial attention weights. These weights are generated based on the contextual guidance of the upsampled decoder features, enabling the network to focus on discriminative regions. Consequently, the refined upsampling and enhanced skip connections allow the decoder to effectively aggregate shallow structural details with deep semantic representations, yielding sharper and more precise segmentation boundaries while suppressing irrelevant background noise.

	\subsection{Auxiliary Supervision}
	To further enhance the training robustness and interpretability of the multi-scale representations, we integrate auxiliary supervision heads (Aux-Head) attached to the intermediate feature maps of the decoder. Specifically, these heads are applied to the outputs at Layer1 and Layer2. 
	Each Aux-Head consists of a lightweight residual convolutional block, which projects the intermediate features to the same number of semantic classes as the main output. The resulting logits are then upsampled to the original spatial resolution via bilinear interpolation to compute the auxiliary segmentation loss. 
	A crucial premise for this design lies in the introduction of the parallel CNN encoder. The CNN's local feature extraction significantly enhances the semantic interpretability of the deep decoder features. Benefiting from this enhanced clarity, even the deeper features processed by the Aux-Head can yield segmentation predictions that are structurally coherent with those from shallower layer, allowing the auxiliary losses to function effectively. 
	It is worth noting that, during our empirical validation, placing the Aux-Heads at Layer3 tends to make the CNN features over-dominant, which actually compromises the overall performance. Therefore, we restrict the auxiliary supervision to Layer1 and Layer2 to ensure a balanced contribution from both encoder branches. In terms of training strategy, we assign a higher loss weight to the auxiliary outputs from shallower layer, while adopting a relatively lower weight for the deeper layer. This weight decay scheme prioritizes the learning of fine-grained structural information from shallow features, while ensuring that the deeper layers also receive gradient guidance, maintaining semantic consistency across the network and boosting the overall interpretability of the intermediate representations.

	\section{Experiments on Synapse Dataset}
	
	\subsection{Dataset and Preprocessing}
	We evaluate our method on the Synapse multi-organ segmentation dataset \cite{landman2015synapse}, a widely used benchmark in medical image analysis. The dataset comprises 30 abdominal CT scans with a total of 3,779 axial slices. Following the standard protocol in prior works, we adopt an 18/12 split, where 18 cases are used for training and the remaining 12 cases for testing. The dataset includes annotations for eight abdominal organs: the aorta, gallbladder, spleen, left kidney, right kidney, liver, pancreas, and stomach. We employ the Dice Similarity Coefficient (DSC) and the 95\% Hausdorff Distance (HD95) as the primary evaluation metrics, which jointly assess the volumetric overlap and the boundary alignment of the predicted segmentations. In the preprocessing stage, all CT slices are resampled to a fixed resolution of $224 \times 224$ via bicubic interpolation. During training, we apply standard augmentations including random $90^\circ$ rotations, flips, and small-angle rotations. Notably, we modify the label resizing pipeline to prevent semantic leakage—a common issue when directly resizing multi-class segmentation masks with global nearest-neighbor interpolation, which can cause thin structures like the pancreas to be erroneously overwritten. We circumvent this by independently resizing per-class binary masks and recombining them using a 0.5 threshold.

	\begin{table}[H]
		\centering
		\caption{Segmentation accuracy of different methods on the Synapse multi-organ CT dataset.}
		\label{tab:results}
		\small
		\begin{tabular}{lccccccccccc}
			\toprule
			Methods & DSC$\uparrow$ & HD95$\downarrow$ & Aor. & Gal. & Kid.L & Kid.R & Liv. & Pan. & Spl. & Sto. \\
			\midrule
			U-Net & 76.85 & 39.70 & 89.07 & 69.72 & 77.77 & 68.60 & 93.43 & 53.98 & 86.67 & 75.58 \\
			nnUNet & \textbf{86.44} & \textit{10.91} & 91.78 & 69.77 & 86.92 & 86.21 & 96.49 & \textit{83.23} & 91.16 & \textit{85.92} \\
			TransUNet & 77.48 & 31.69 & 87.23 & 63.13 & 81.87 & 77.02 & 94.08 & 55.86 & 85.08 & 75.62 \\
			Swin-Unet & 79.13 & 21.55 & 85.47 & 66.53 & 83.28 & 79.61 & 94.29 & 56.58 & 90.66 & 76.60 \\
			Swin-UNETR & \textbf{83.48} & \textit{10.55} & 91.12 & 66.54 & 86.99 & 86.26 & 95.72 & 68.80 & 95.37 & 77.01 \\
			nnFormer & 86.57 & \textit{10.63} & 92.04 & 70.17 & 86.57 & 86.25 & 96.84 & \textit{83.35} & 90.51 & \textit{86.83} \\
			MISSformer & 81.96 & 18.20 & 86.99 & 68.65 & 85.21 & 82.00 & 94.41 & 65.67 & 91.92 & 80.81 \\
			HiFormer-L & 80.69 & 19.14 & 87.03 & 68.61 & 84.23 & 78.37 & 94.07 & 60.77 & 90.44 & 82.03 \\
			SAM3D & 79.56 & 17.87 & 89.57 & 49.81 & 86.31 & 85.64 & 95.42 & 69.32 & 84.29 & 76.11 \\
			UNETR++ & \textbf{87.22} & \textit{7.53} & 92.52 & 71.25 & 87.54 & 87.18 & 96.42 & \textit{81.10} & 95.77 & \textit{86.01} \\
			VM-UNet & 81.08 & 19.21 & 86.40 & 69.41 & 86.16 & 82.76 & 93.63 & 58.36 & 89.51 & 81.40 \\
			CiUNet & \textbf{83.72} & 12.13 & 88.40 & \textbf{79.47} & \textbf{91.39} & \textbf{91.27} & 93.61 & 59.31 & 90.33 & 76.00 \\
			\bottomrule
		\end{tabular}
	\end{table}
	
	\subsection{Implementation Details}
	Our method is implemented using Python 3.7 and the PyTorch framework (v1.13.1, CUDA 11.6), and trained on an NVIDIA A2000 GPU with 12GB memory. The Swin Transformer encoder and the shallow texture extraction layers of the CNN encoder are initialized with pre-trained weights. The initial learning rate is set to the baseline of Swin-Unet \cite{cao2021swinunet}, with a weight decay dynamically adjusted from \(1 \times 10^{-5}\) to \(1 \times 10^{-4}\) along with the training progress aiming for completion between 150 and 300 epochs. We adopt the SGD optimizer with a momentum of 0.9. All input images and labels are resized to $224 \times 224$ during both training and testing. It should be noted that all quantitative metrics for our model reported in the subsequent tables are evaluated by upsampling the $224 \times 224$ predictions back to the original $512 \times 512$ resolution, whereas the provided demo implementation operates natively at the $224 \times 224$ resolution.
	\clearpage
	
	The overall optimization objective is formulated as a weighted combination of three distinct geometric constraints: pixel-level classification, region-level overlap, and boundary alignment. The main loss $\mathcal{L}_{\text{main}}$ is defined as:
	
	\begin{equation}
		\mathcal{L}_{\text{main}} = \sum_{i \in \{\text{ce}, \text{dc}, \text{hd}\}} \left( \lambda_i \cdot \mathcal{G}_i \cdot \mathcal{L}_i \right)
	\end{equation}
	where $\mathcal{L}_i$ denotes the individual loss components. \\\\
	For the CE counterpart, we employ the Focal Loss implemented in the Kornia library \cite{eriba2019kornia}, with $\alpha=0.25$ and $\gamma=2.0$ to effectively mitigate the severe class imbalance in multi-organ segmentation. 
	\begin{equation}
		\mathcal{L}_{\text{ce}} = -\alpha_t (1 - p_t)^{\gamma} \log(p_t) \quad \text{\cite{lin2018focal}}
	\end{equation}
	For the DC counterpart, we adopt a dynamic weighting strategy that progressively increases the boundary weights in the Dice loss as training advances, thereby adaptively prioritizing boundary regions and improving the overlap consistency for small anatomical structures. Specifically, we generate a pixel-wise boundary weight map \(w_b = 1 - (1 - \mathbf{p})^{\beta}\), where \(\mathbf{p}\) denotes the output probability map and \(\beta\) controls the attenuation rate of the boundary focus. This weight map is dynamically integrated into the Dice formulation to effectively amplify gradients along organ contours while suppressing the influence of the homogeneous interior. The weighted DC loss is formally defined as:
	\begin{equation}
		\mathcal{L}_{\text{dc}} = 1 - \frac{2 \sum_{i} \left( w_b \cdot p_i \cdot g_i \right) + \epsilon}{\sum_{i} \left( w_b \cdot p_i \right) + \sum_{i} \left( w_b \cdot g_i \right) + \epsilon}
	\end{equation}
	where \(p_i\) and \(g_i\) denote the predicted probability and ground-truth label at pixel \(i\), respectively, and \(\epsilon\) is a small constant for numerical stability.\\\\
	For the HD counterpart, employ the HausdorffERLoss implemented in the Kornia library, which approximates the Hausdorff distance via morphological erosion. 
	\begin{equation}
		\mathcal{L}_{\text{hd}} = \frac{1}{|\Omega|} \sum_{k=1}^{K} \sum_{\Omega} \left( (p - q)^2 \ominus_k B \right) \cdot k^{\alpha} \quad \text{\cite{karimi2019reducing}}
	\end{equation}

	The hyperparameters of this loss are dynamically scheduled along the training curriculum: the erosion exponent \(\alpha\) is gradually increased from \(1.0\) to \(2.0\), while the number of erosion iterations \(k\) is progressively decreased from approximately \(15\%\) of the image size (\(224\) pixels) down to \(5\%\) (11 iterations). In our implementation, the balancing weights and gain factors are empirically determined based on the numerical behaviors documented in the Kornia documentation and our experimental feedback. Specifically, we set the loss weights to \(\lambda_{\text{ce}}=0.3\), \(\lambda_{\text{dc}}=0.6\), \(\lambda_{\text{hd}}=0.1\), and the corresponding gain factors to \(\mathcal{G}_{\text{ce}}=1.0\), \(\mathcal{G}_{\text{dc}}=1.0\), and \(\mathcal{G}_{\text{hd}}=10.0\). This configuration ensures that the contributions from the CE and DC terms remain well-conditioned during training, while the inherently small magnitude of the HausdorffER Loss is appropriately amplified without disrupting the overall loss balance.
	
	In addition to the main segmentation loss, we integrate auxiliary supervision heads (Aux-Head) attached to the intermediate decoder feature maps. This design serves as an exploratory attempt to visualize deep semantic representations, further encouraging the network to produce structurally coherent predictions across different feature levels. Each head adopts a lightweight residual structure comprising a 3\(\times\)3 Convolution, GroupNorm, GELU, and a 1\(\times\)1 Convolution for channel alignment, with a shortcut connection analogous to a standard ResNet block.
	
	\clearpage
	
	\subsection{Curriculum Learning Strategy}
	To facilitate robust training and prevent early overfitting, we adopt a multi-stage curriculum learning strategy that gradually exposes the network to increasingly diverse and challenging training data. We design 11 progressive stages, each associated with a complexity level ranging from 0 to 1.0 at intervals of 0.1. The core principle lies in the progressive enrichment of the training set composition. At the initial training stages (lower complexity levels), the training data is predominantly curated to include only slices containing the focal organs (pancreas and gallbladder). As the complexity level increases, the data pool expands by progressively incorporating more background-dominant slices and slices containing other abdominal organs. By the final stage (complexity level 1.0), the training set encompasses the full distribution of available non-empty slices, exposing the model to the complete anatomical variability present in the dataset. Specifically, for each complexity level, a dedicated training and validation split is dynamically generated using our custom data partitioning algorithm. At each stage, the model is continuously trained until a new best checkpoint is achieved, with a minimum training duration of \(3 + \text{stage\_index}\) epochs. Once the model reaches the predefined epoch target and successfully updates the best checkpoint, the training for the current stage is completed. The optimal model from the preceding stage serves as the initialization for the subsequent stage, enabling smooth knowledge transfer across the curriculum. After completing all 11 stages, the model continues its training until 300 epochs on the final full dataset (complexity level 1.0), achieving the highest segmentation performance. This progressive paradigm effectively mitigates the learning burden in early stages; during this process, the validation loss does not drop as rapidly as it does in the original Swin-UNet \cite{cao2021swinunet}, which helps build a solid foundation rather than pursuing quick gains.
	\begin{figure}[htbp]
		\centering
		\includegraphics[width=1.00\linewidth, keepaspectratio]{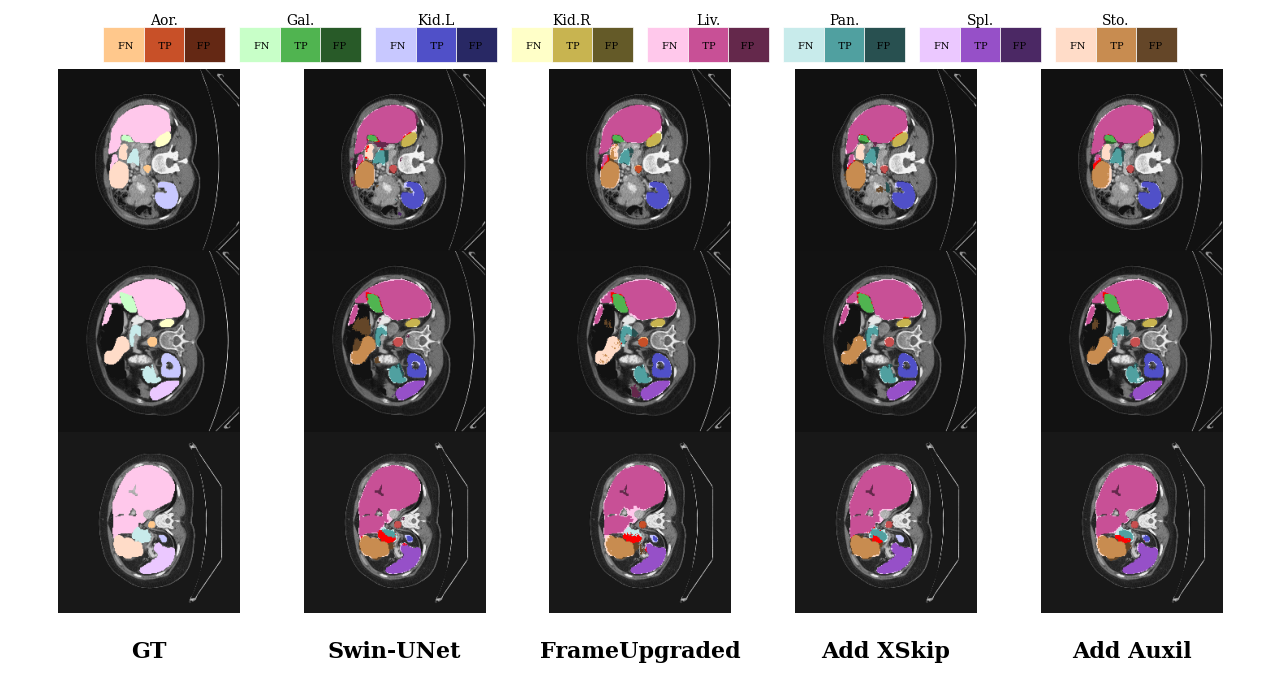}
		\caption{The segments of different modification stages on the Synapse dataset.}
		\label{fig:ablation}
	\end{figure}
	
	\subsection{Experiment Results}
	As shown in Table \ref{tab:results}, CiUNet achieves a mean DSC of 83.72\% and a mean HD95 of 12.13 mm on the Synapse dataset. It should be noted that in the table, bold entries indicate superior DSC performance, whereas italicized entries denote fine boundary precision (HD95) typically achieved by either 3D segmentation models or models that operate natively at $512 \times 512$. 
	\clearpage
	In terms of overall segmentation accuracy, our model surpasses the original Swin-Unet (79.13\%), TransUNet (77.48\%), and MISSformer (81.96\%), while remaining comparable to stronger baselines such as Swin-UNETR (83.48\%) and HiFormer-L (80.69\%). 
	When examining the per-organ results, CiUNet exhibits particularly strong performance on the gallbladder and both kidneys, achieving the highest DSC scores across all compared methods (Gallbladder: 79.47\%; Left Kidney: 91.39\%; Right Kidney: 91.27\%). This indicates that the dual-encoder design and the XSkip mechanism are highly effective in preserving high-resolution spatial details for relatively well-defined organs. However, the performance on the pancreas (59.31\%) and stomach (76.00\%) remains suboptimal compared to the state-of-the-art methods like nnFormer and UNETR++.
	It is worth noting that the results reported for CiUNet are obtained from the checkpoint at epoch 204, where the auxiliary supervision strategy is fully employed.
	
	\begin{table}[htbp]
		\caption{Stepwise ablation study on the individual contributions of the dual-encoder, skip connections, and auxiliary supervision.}
		\label{tab:ablation}
		\centering
		\resizebox{\textwidth}{!}{%
			\begin{tabular}{l|cc|cc|cccccccc}
				\hline
				\multirow{2}{*}{\textbf{}} & \textbf{Params} & \textbf{Epoch} & \textbf{DSC\(\uparrow\)} & \textbf{HD95\(\downarrow\)} & \textbf{Aor.} & \textbf{Gal.} & \textbf{Kid.L} & \textbf{Kid.R} & \textbf{Liv.} & \textbf{Pan.} & \textbf{Spl.} & \textbf{Sto.} \\
				\hline
				Swin-Unet Baseline & 27.17M & 150 & 79.13 & 21.55 & 85.47 & 66.53 & 83.28 & 79.61 & 94.29 & 56.58 & 90.66 & 76.60 \\
				\hline
				+ Framework Upgrades & 30.22M & 150 & 81.40 & 15.43 & 86.15 & 75.40 & 87.79 & 87.03 & 93.53 & 54.47 & 91.66 & 75.20 \\
				& & 280 & 81.47 & 14.68 & 86.53 & 76.43 & 87.85 & 86.68 & 93.68 & 55.05 & 91.25 & 74.31 \\
				\hline
				+ XSkip & 31.68M & 161 & 83.18 & 15.59 & 88.00 & 77.40 & 88.88 & 89.49 & 93.56 & 61.63 & 90.94 & 75.53 \\
				& & 273 & 83.54 & 11.26 & 88.44 & 73.61 & 89.71 & 91.11 & 93.77 & 62.15 & 92.01 & 77.50 \\
				\hline
				+ Aux & 31.68M & 148 & 83.29 & 13.21 & 87.58 & 77.68 & 90.42 & 90.86 & 93.20 & 62.59 & 88.75 & 75.26 \\
				& & 204 & 83.72 & 12.13 & 88.40 & 79.47 & 91.39 & 91.27 & 93.61 & 59.31 & 90.33 & 76.00 \\
				\hline
			\end{tabular}%
		}
	\end{table}

	\subsection{Ablation Study}
	To systematically evaluate the contribution of each proposed component, we conduct ablation experiments on the Synapse test set. All models are inferred at \(224 \times 224\) and evaluated at the original \(512 \times 512\) resolution to ensure fair metric comparison. The results are summarized in Table.\ref{tab:ablation} and Fig.\ref{fig:ablation}
	
	\textbf{Effect of Framework Upgrades:} The Swin-Unet baseline reaches 79.13\% DSC and 21.55 HD95 with 27.17M parameters at 150 epochs. To capture both global context and local spatial details, we introduce a dual-encoder architecture incorporating a Swin Transformer branch and a parallel CNN branch. In this upgraded framework, we also integrate the progressive curriculum learning strategy and the multi-component loss design as foundational components rather than isolated modifications. At 150 epochs, this consolidated framework already achieves 81.40\% DSC and 15.43 HD95, demonstrating significant improvements in both overlap and boundary alignment even at an early convergence point. Extending the training to 280 epochs yields marginal gains (81.47\% DSC, 14.68 HD95), indicating the structural upgrades themselves enable faster convergence over the baseline.
	
	\textbf{Effect of XSkip Connections:} Building on the upgraded framework, we introduce enhanced XSkip pathways with spatial alignment and channel compression. This configuration dramatically improves boundary alignment. At 161 epochs, the model with XSkip reaches 83.18\% DSC and 15.59 HD95; further training to 273 epochs pushes the HD95 to 11.26 while the DSC rises to 83.54\%. The sharp drop in HD95 verifies that our skip connections are highly effective at preserving high-resolution spatial details lost during aggressive down-sampling.
	
	\textbf{Effect of Auxiliary Supervision:} Finally, we add Aux-Heads with customized loss weighting. Benefiting from the CNN features that enhance deep semantic interpretability, this mechanism significantly accelerates training convergence. At only 148 epochs, the model already achieves 83.29\% DSC and 13.21mm HD95. The optimal performance is reached at 204 epochs, yielding a peak DSC of 83.72\% with an HD95 of 12.13mm. This confirms that auxiliary supervision effectively mimics human cognitive learning via multi-level gradient guidance, simultaneously improving convergence speed and final segmentation accuracy.
	\clearpage
	
	\section{Discussion}
	While CiUNet demonstrates competitive performance on the Synapse dataset, several limitations and promising directions for future improvement should be acknowledged.
	
	\textbf{Metric bias toward false positives:} The evaluation metrics commonly used in medical segmentation inherently encourage false positives. Specifically, the Dice coefficient and HD95 tend to reward over-segmentation, as extending the predicted boundaries slightly outward often improves overlap without significantly penalizing the Hausdorff distance. This bias suggests that incorporating explicit false positive suppression mechanisms into the loss design would be a valuable direction for further refinement.
	
	\textbf{Sensitivity to boundary slices in 3D anatomy:} The performance gap between 2D and 3D models on small organs—such as the pancreas and gallbladder—remains non-negligible. The superior results achieved by 3D models like UNETR++ and nnFormer largely benefit from volumetric continuity across adjacent slices. In 2D slice-wise segmentation, this inter-slice contextual information cannot be adequately compensated by simply propagating shallow features to deeper layers. Furthermore, small organs are particularly sensitive to the start and end frames of the volumetric sequence, where partial-volume effects and ambiguous boundaries introduce additional uncertainty. Therefore, developing lightweight mechanisms to incorporate inter-slice awareness within a 2D framework represents a meaningful future challenge.
	
	\textbf{Suboptimal backbone initialization:} The current backbone design, which adopts Swin-Tiny for global semantic reasoning and a ResNet-based parallel branch for texture extraction, serves as a reliable starting point but is inherently suboptimal. Neither backbone is specifically tailored for medical image characteristics. The Swin-Tiny weights are transferred from natural image pre-training, while the ResNet texture features are only shallowly frozen. A more principled approach would involve a dedicated pre-training strategy jointly optimized for both branches on in-domain medical data.
	
	\section{Conclusion}
	In this paper, we have presented CiUNet, an improved variant built upon the Swin-UNet architecture for medical image segmentation. By introducing a hybrid framework that synergistically integrates the global reasoning capabilities of Swin Transformer with the local texture sensitivity of a parallel CNN, our model effectively alleviates the inherent limitations of pure Transformer-based U-shaped networks. Through the incorporation of a cross-branch fusion module and XSkip connections, the proposed architecture mitigates the spatial information loss introduced by aggressive down-sampling. Additionally, the auxiliary supervision heads serve as a valuable guide for deep feature learning, promoting more interpretable and structurally coherent intermediate representations. Extensive experiments on the Synapse multi-organ segmentation dataset demonstrate that CiUNet achieves competitive performance with significantly fewer trainable parameters compared to existing 3D methods, striking a favorable balance between segmentation accuracy and computational efficiency. 
	\clearpage
	
	\bibliographystyle{plainnat}
	\bibliography{references}
	
\end{document}